\documentclass[aps,prl, nofootinbib,twocolumn]{revtex4}

\usepackage{epsfig}
\usepackage{color}
\usepackage[latin1]{inputenc}
\usepackage{float,amsmath}
\usepackage{graphicx}

\newcommand{\be}{\begin{eqnarray}}
\newcommand{\ee}{\end{eqnarray}}

\newcommand{\benum}{\begin{enumerate}}
\newcommand{\eenum}{\end{enumerate}}

\begin{document}

\title{Elliptic and triangular flow in event-by-event (3+1)D viscous hydrodynamics}

\author{Bj\"orn Schenke}
\affiliation{Department of Physics, McGill University, 3600 University Street, Montreal, Quebec, H3A\,2T8, Canada}

\author{Sangyong Jeon}
\affiliation{Department of Physics, McGill University, 3600 University Street, Montreal, Quebec, H3A\,2T8, Canada}

\author{Charles Gale}
\affiliation{Department of Physics, McGill University, 3600 University Street, Montreal, Quebec, H3A\,2T8, Canada}

\begin{abstract}
We present results for the elliptic and triangular flow coefficients $v_2$ and $v_3$ in Au+Au collisions at $\sqrt{s}=200A\,{\rm GeV}$ 
using event-by-event 3+1D viscous hydrodynamic simulations. We study the effect of initial state fluctuations and finite viscosities
on the flow coefficients $v_2$ and $v_3$ as functions of transverse momentum and pseudo-rapidity.
Fluctuations are essential to reproduce the measured centrality dependence of elliptic flow.
We argue that simultaneous measurements of $v_2$ and $v_3$ can determine $\eta/s$ more precisely.
\end{abstract}

\maketitle


Fluctuating initial conditions for hydrodynamic simulations of heavy-ion collisions
have been argued to be very important for the exact determination of collective flow observables and to describe 
specific features of multi-particle correlation measurements in heavy-ion collisions at Brookhaven National Laboratory's
 Relativistic Heavy-Ion Collider (RHIC)
\cite{Adare:2008cqb,Abelev:2008nda,Alver:2008gk,Alver:2009id,Abelev:2009qa,Miller:2003kd,Broniowski:2007ft,Hirano:2009bd,
Takahashi:2009na,Andrade:2009em,Alver:2010gr,Werner:2010aa,Holopainen:2010gz,Alver:2010dn,Petersen:2010cw}.
Both long-range correlations in pseudo-rapidity and double-peak structures on the away-side in $\Delta\eta_p$-$\Delta\phi$-correlations 
have been reproduced using initial states with fluctuations in the transverse plane, extending as flux-tubes along the beam-line 
\cite{Takahashi:2009na,Andrade:2009em,Alver:2010gr,Alver:2010dn}.

In this work we report on results for both elliptic and triangular flow obtained with 
an event-by-event 3+1D relativistic hydrodynamic simulation, an extension of 
\textsc{music} \cite{Schenke:2010nt}, including shear viscosity. 
We first briefly describe the inclusion of viscosity and leave a more detailed description to a forthcoming work.

In the first order formalism for viscous hydrodynamics,
the stress-energy tensor is decomposed into
\be
T_{\rm 1st}^{\mu\nu} = T^{\mu\nu}_{\rm id} + S^{\mu\nu}\,,
\ee
where
\be
T_{\rm id}^{\mu\nu} = (\epsilon + {\cal P})u^\mu u^\nu - {\cal P}g^{\mu\nu}
\ee
is the ideal fluid part with flow velocity $u^\mu$, local energy density
$\epsilon$ and local pressure ${\cal P}$. The flow velocity is defined as
the time-like eigenvector of $T^{\mu\nu}_{\rm id}$
\be
T^{\mu\nu}_{\rm id} u_{\nu} = -\epsilon u^\nu
\ee
with the normalization $u^\nu u_\nu = 1$ and the pressure is determined by
the equation of state as a function of $\epsilon$.

The viscous part of the stress energy tensor in the first-order approach is
given by 
\be
S^{\mu\nu} = \eta 
\left(
\nabla^\mu u^\nu + \nabla^\nu u^\mu - 
{2\over 3}\Delta^{\mu\nu}\nabla_\alpha u^\alpha
\right)
\ee
where $\Delta^{\mu\nu} = g^{\mu\nu} - u^\mu u^\nu$ is the local 3-metric 
and $\nabla^\mu = \Delta^{\mu\nu}\partial_\nu$ is the local space derivative.
Note that $S^{\mu\nu}$ is transverse with respect to the flow velocity since
$\Delta^{\mu\nu}u_\nu = 0$ and $u^\nu u_\nu = 1$. 
Hence, $u^\mu$ is also an
eigenvector of the whole stress-energy tensor with the same eigenvalue
$\epsilon$.

This form of viscous hydrodynamics is conceptually simple. However, this Navier-Stokes form is known to introduce unphysical 
superluminal signals. There are several remedies for this
problem 
\cite{Israel:1976tn,Stewart:1977,Israel:1979wp,Grmela:1997zz,Muronga:2001zk}, 
all of them employing the second order formalism. 
In this work, we use a variant of the Israel-Stewart formalism
derived in \cite{Baier:2007ix}, where the stress-energy tensor is decomposed as
\be\label{tmunu}
{\cal T}^{\mu\nu} = T^{\mu\nu}_{\rm id} + W^{\mu\nu}\,.
\ee
The evolution equations are
$\partial_\mu {\cal T}^{\mu\nu} = 0$
and
\begin{equation}
\Delta^{\mu}_{\alpha}\Delta^{\nu}_{\beta}
{u^\sigma\partial_\sigma} W^{\alpha\beta}
=
-{1\over \tau_\pi}
\left( W^{\mu\nu} - S^{\mu\nu} \right) - {4\over 3}W^{\mu\nu}(\partial_\alpha u^\alpha)\,.
\label{eq:Weq}
\end{equation}

In the $\tau, \eta_s$ coordinate system we use, these equations can be
re-written as hyperbolic equations with sources
\be
\partial_a T_{\rm id}^{ab} = -\partial_a W^{ab} + F^b
\label{eq:Tid_eq}
\ee
and
\be
\partial_a (u^a W^{cd}) = -(1/\tau_\pi)(W^{cd} - S^{cd}) + G^{cd}
\label{eq:uW_eq}
\ee
where $F^b$ and $G^{cd}$ contain terms introduced by the coordinate change
from $t,z$ to $\tau,\eta_s$ as well as those introduced by the projections in
Eq.(\ref{eq:Weq}).

Our approach to solve these hyperbolic equations relies on
the Kurganov-Tadmor (KT) scheme
\cite{Kurganov:2000,Naidoo:2004}, together
with Heun's method to solve resulting ordinary
differential equations. For details, see \cite{Schenke:2010nt}.
The main difference between the method employed here and the method 
used to solve
ideal hydrodynamics in \cite{Schenke:2010nt} is the appearance of
time-derivatives in the source term. These are handled with the first order
approximation
$\dot{g}(\tau_n) = (g(\tau_{n})-g(\tau_{n-1}))/\Delta \tau$
in the first step of the Heun method,
and in the second step we use
$\dot{g}(\tau_n) = (g^*(\tau_{n+1})-g(\tau_n))/\Delta \tau$
where $g^*(\tau_{n+1})$ is the result from the first step.

As in most Eulerian algorithms, ours also suffers from numerical
instability when the density becomes small while the flow velocity becomes
large. Fortunately this happens late in the evolution of the system.
Regularizing such instability has no strong effects on the observables we
are interested in. 
Some ways of handling this are known (for instance see Ref.\cite{Duez:2004nf}). 
\begin{widetext}

\begin{figure}[t]
\vspace{-0.5cm}
   \begin{center}
     \includegraphics[width=18cm]{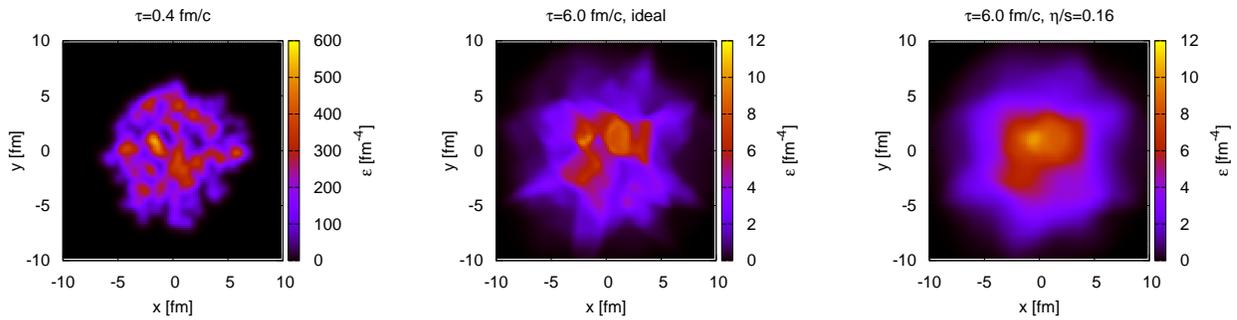}
\vspace{-0.75cm}
     \caption{(Color online) Energy density distribution in the transverse plane for one event with $b=2.4\,{\rm fm}$
       at the initial time (left), and after $\tau=6\,{\rm fm}/c$ for the ideal case (middle) and with $\eta/s=0.16$ (right).\label{fig:three}}
   \end{center}
\vspace{-0.75cm}
\end{figure}
\end{widetext}
In this study, we found that setting the local viscosity to zero 
when finite viscosity causes negative pressure in the cell as advocated 
in \cite{Duez:2004nf} and reducing the ideal part by 5\,\% works well to
stabilize the calculations without introducing spurious effects.

While in standard hydrodynamic simulations with averaged initial conditions all odd flow coefficients
vanish by definition, fluctuations generate triangular flow $v_3$ as a response to the finite initial triangularity.

We follow \cite{Petersen:2010cw} and define
an event plane through the angle
\begin{equation}
  \psi_n=\frac{1}{n}\arctan\frac{\langle p_T \sin(n\phi)\rangle}{\langle p_T \cos(n\phi)\rangle}\,,
\end{equation}
where the weight $p_T$ is chosen for best accuracy \cite{Poskanzer:1998yz}.
Then, the flow coefficients can be computed using
\begin{equation}
  v_n=\langle \cos(n(\phi-\psi_n)) \rangle\,.
\end{equation}

The initialization of the energy density is done using a Glauber Monte-Carlo model (see \cite{Miller:2007ri}):
Before the collision the density distribution of the two nuclei is described by
a Woods-Saxon parametrization, which we sample to determine the positions of individual nucleons.
The impact parameter is sampled from the distribution 
$P(b) db = 2bdb/(b_{\rm max}^2-b_{\rm min}^2)$, where
$b_{\rm min}$ and $b_{\rm max}$ depend on the given centrality class.
Then we determine the distribution of binary collisions and wounded nucleons. 
Two nucleons are assumed to collide if their relative transverse distance is less than
$ D = \sqrt{\sigma_{NN}/\pi}$,
where $\sigma_{NN}$ is the inelastic nucleon-nucleon cross-section, which at top RHIC energy of $\sqrt{s}=200A\,{\rm GeV}$ is $\sigma_{NN}=42\,{\rm mb}$.
The energy density is distributed proportionally to the wounded nucleon distribution. 
For every wounded nucleon we add a contribution to the energy density with Gaussian shape (in $x$ and $y$) and width $\sigma_0=0.4\,{\rm fm}$.
In the rapidity direction, we assume the energy density to be constant on a central plateau and fall like half-Gaussians at large $|\eta_s|$ 
(see \cite{Schenke:2010nt}).
This procedure generates flux-tube like structures compatible with measured long-range rapidity correlations \cite{Jacobs:2005pk,Wang:2004kfa,Adams:2005ph}.
The absolute normalization is determined by demanding that the obtained total multiplicity distribution reproduces the experimental data.

As equation of state we employ the parametrization ``s95p-v1'' from \cite{Huovinen:2009yb}, obtained from interpolating between
lattice data and a hadron resonance gas.

In Fig.\,\ref{fig:three} we show the energy density distribution in the transverse plane 
for an event with impact parameter $b=2.4\,{\rm fm}$ at the initial time $\tau_0=0.4\,{\rm fm}/c$
and at time $\tau=6\,{\rm fm}/c$ for $\eta/s=0$ and $\eta/s=0.16$.
This clearly shows the effect of dissipation. 

We perform a Cooper-Frye freeze-out using
\begin{equation}\label{cf}
E\frac{dN}{d^3p}=\frac{dN}{dy p_T dp_T d\phi_p} = g_i \int_\Sigma f(u^\mu p_\mu) p^\mu d^3\Sigma_\mu\,,
\end{equation}
where $g_i$ is the degeneracy of particle species $i$, and $\Sigma$ the freeze-out hyper-surface.
In the ideal case the distribution function is given by
\begin{equation}
  f(u^\mu p_\mu) = f_0(u^\mu p_\mu) = \frac{1}{(2\pi)^3}\frac{1}{\exp((u^\mu p_\mu -\mu_i)/T_{\rm FO})\pm 1}\,,
\end{equation}
where $\mu_i$ is the chemical potential for particle species $i$ and $T_{\rm FO}$ is the freeze-out temperature.
In the finite viscosity case we include viscous corrections to the distribution function, $f=f_0+\delta f$, with
\begin{equation}
 \delta f = f_0 (1\pm f_0) p^\alpha p^\beta W_{\alpha\beta} \frac{1}{2 (\epsilon+\mathcal{P}) T^2}\,,
\end{equation}
where $W$ is the viscous correction introduced in Eq. (\ref{tmunu}).
Note that the choice $\delta f \sim p^2$ is not unique \cite{Dusling:2009df}.

\begin{figure}[tb]
   \begin{center}
\vspace{-0.4cm}
     \includegraphics[width=8.5cm]{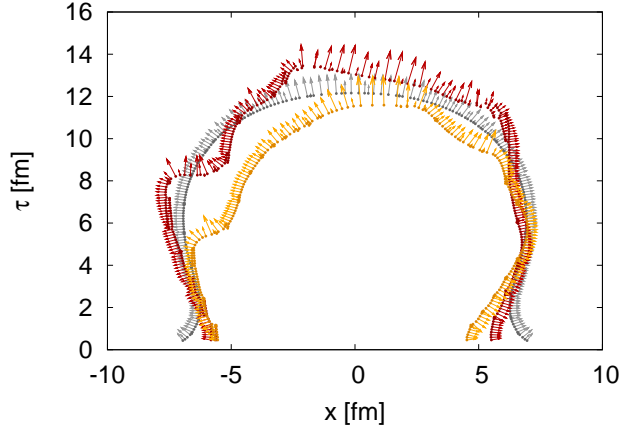}
\vspace{-0.3cm}
     \caption{(Color online) Freeze-out surfaces for two different events
      (red and yellow) compared to that for the averaged initial condition (gray).}
     \label{fig:compareFO}
   \end{center}
\vspace{-0.5cm}
\end{figure}

The algorithm used to determine the freeze-out surface $\Sigma$ has been presented in \cite{Schenke:2010nt}. 
It is very efficient in determining the freeze-out surface of a system with fluctuating initial conditions. To demonstrate this, we
present the freeze-out surface in the $x$-$\tau$-plane in the vicinity of $y=0\,{\rm fm}$ and $\eta_s=0$ for two different initial distributions compared to that for an averaged initial condition in Fig.\,\ref{fig:compareFO}. The arrows are projections 
of the normal vector on the hyper-surface element onto the $x$-$\tau$ plane.

We include resonances up to the $\phi$-meson. We found that the pseudorapidity dependence of both $v_2$ and $v_3$ is affected notably by the
inclusion of resonance decays, improving the agreement of $v_2(\eta_p)$ with data significantly. $v_2(p_T)$ at midrapidity is almost unaffected by the resonances while $v_3(p_T)$ is reduced by approximately 20-30\%.

\begin{figure}[tb]
   \begin{center}
     \includegraphics[width=8.5cm]{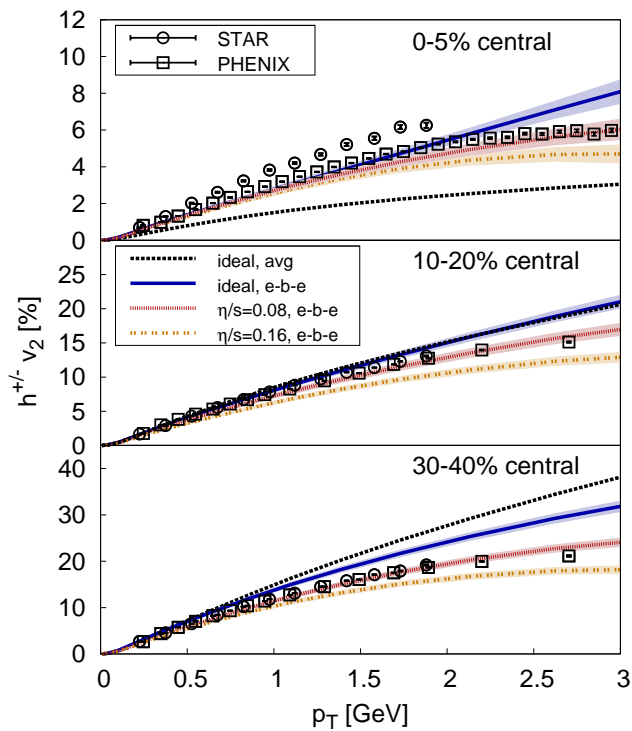}
\vspace{-0.5cm}
     \caption{(Color online) Charged hadron $v_2$ for different centralities as a function of transverse momentum for averaged 
       initial conditions (avg) and event-by-event simulations (e-b-e) using
       different viscosity to entropy density ratios compared to STAR \cite{Adams:2004bi} and PHENIX \cite{Adare:2010ux} data.}
     \label{fig:v2hc}
   \end{center}
\vspace{-0.75cm}
\end{figure}

Fig.\,\ref{fig:v2hc} shows the elliptic flow $v_2$ for charged hadrons as a
function of transverse momentum obtained from an averaged initial condition 
in the ideal case and for an average over 100 individual events for $\eta/s \in \{0,0.08,0.16\}$.
We compare to data from STAR \cite{Adams:2004bi} and PHENIX \cite{Adare:2010ux}. The used minimal, maximal, and average
impact parameter for each centrality class are given in Table \ref{table1} .
\begin{table}[tb]
\centering
\begin{tabular}{|c|c|c|c|}
\hline
centrality [\%]	& $b_{\rm min}$ [fm] & $b_{\rm max}$ [fm] & $\langle b\rangle$ [fm]\\ 
\hline
0-5	& 0     &  3.37  &  2.24  \\
10-20   & 4.75  &  6.73  &  5.78  \\
15-25   & 5.83  &  7.53  &  6.7   \\
30-40	& 8.23  &  9.5   &  8.87  \\
\hline
\end{tabular}
\caption{\protect\small Used impact parameters. \label{table1}} 
\vspace{-0.5cm}
\end{table}
While in the most central collisions fluctuations increase $v_2$ compared to
the case with averaged initial conditions, for 
10-20\% central collisions the difference is negligible and for 30-40\% central collisions fluctuations reduce the elliptic flow.
The increase for central collisions is easy to understand since we are now determining $v_2$ in every event.
Single events have a larger anisotropy with respect to the event-plane than the average with respect to the reaction plane, hence
increasing the obtained $v_2$.
This effect decreases with increasing centrality eventually making the
event-by-event $v_2$ smaller compared to the averaged initial condition case.
This can be understood by the fact that for more peripheral collisions, 
lumps in the initial condition tend not to align perfectly with the
statistically determined event plane.

Viscosity reduces the elliptic flow for all centralities 
as also found in (2+1)-dimensional simulations \cite{Romatschke:2007mq,Dusling:2007gi,Molnar:2008xj,Song:2008si}.

\begin{figure}[tb]
   \begin{center}
     \includegraphics[width=8.5cm]{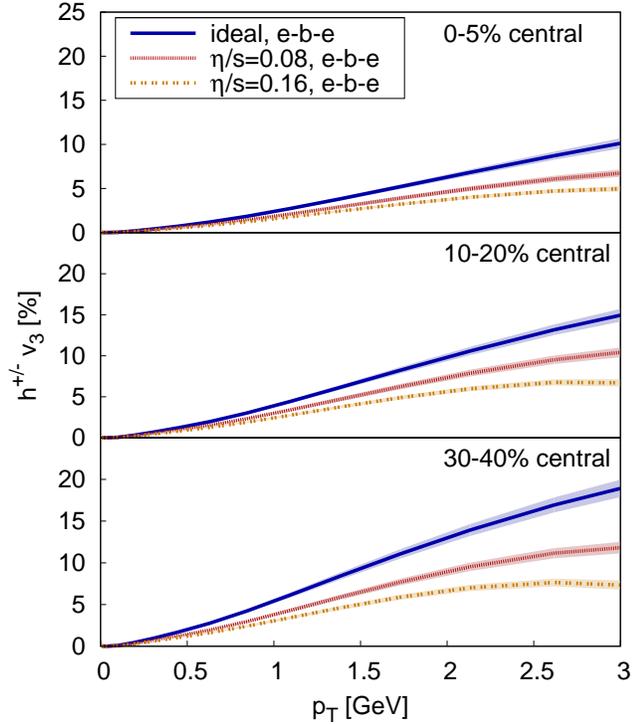}
\vspace{-0.5cm}
     \caption{(Color online) Charged hadron $v_3$ for different centralities as a function of $p_T$ for event-by-event simulations using
       different viscosity to entropy density ratios.}
     \label{fig:v3hc}
   \end{center}
\vspace{-0.5cm}
\end{figure}

Triangular flow $v_3$ as a function of transverse momentum is shown in Fig.\,\ref{fig:v3hc}.
$v_3$ depends less strongly on the centrality than $v_2$ since it is completely fluctuation driven. 
It is largest for an ideal fluid and reduces similarly to $v_2$ with increasing viscosity of the medium.

\begin{figure}[tb]
   \begin{center}
     \includegraphics[width=8.5cm]{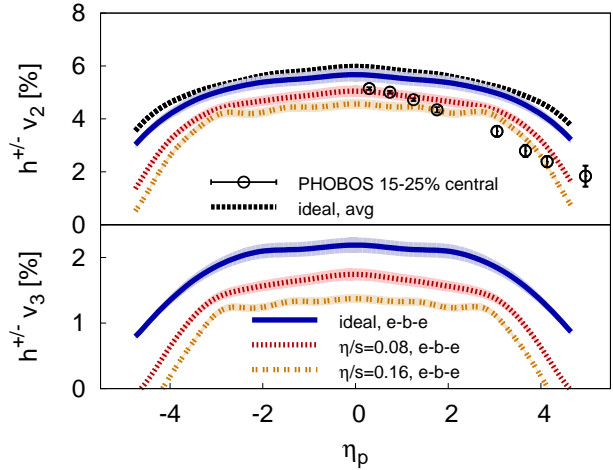}
     \vspace{-0.5cm}
     \caption{(Color online) Charged hadron $v_2$ and $v_3$ as a function of pseudo-rapidity for event-by-event simulations using
       different viscosity to entropy density ratios.}
     \label{fig:v2v3etahc}
   \end{center}
\vspace{-0.9cm}
\end{figure}

The upper panel of Fig.\,\ref{fig:v2v3etahc} shows the pseudo-rapidity dependence of $v_2$ for 15-25\% central collisions 
compared to PHOBOS data \cite{Back:2004mh}.
A reduction of elliptic flow with increasing viscosity is visible, particularly for large pseudo-rapidities $|\eta_p|$, which has been anticipated 
\cite{Bozek:2009mz,Schenke:2010nt}.
In the lower panel of Fig.\,\ref{fig:v2v3etahc} we present the $\eta_p$-dependence of $v_3$. 
Again, the decrease of $v_3$ with increasing viscosity is visible, being strongest for large $|\eta_p|$.

We presented the elliptic and triangular flow coefficients obtained with an event-by-event analysis using (3+1)-dimensional
relativistic viscous hydrodynamics. Charged hadron elliptic flow around midrapidity is well described for a wide range of centralities
when using $\eta/s=0.08$, the conjectured lower bound from AdS/CFT \cite{Kovtun:2004de}. A similarly small value was found in a parton cascade model
based on perturbative QCD \cite{Xu:2007jv}. Larger viscosities underestimate elliptic flow. 
Shear viscosity reduces $v_2$ especially for larger pseudo-rapidities, however, the data is still overestimated away from midrapidity.
Triangular flow has a weaker dependence on centrality. We determined its transverse momentum and pseudo-rapidity dependence, as well as its dependence on $\eta/s$. When triangular flow data becomes available, combined analyses of both $v_2$ and $v_3$ can make an accurate determination of the 
shear-viscosity possible.

\emph{Acknowledgments}
We thank Pasi Huovinen and Paul Romatschke for helpful comments and Todd Springer for fruitful discussions and checks.
This work was supported in part by the Natural Sciences and Engineering Research Council of Canada. B. S.\ gratefully acknow\-ledges a Richard H.~Tomlinson Postdoctoral Fellowship by McGill University.

\bibliography{hydro}

\end{document}